\documentclass[journal,twocolumn,10pt,twoside]{IEEEtranTCOM}

\usepackage{amsthm}
\usepackage{amsmath}
\usepackage{amssymb}
\usepackage{subfigure}
\usepackage{graphicx}
\usepackage{cite}
\usepackage{subfigure}
\usepackage{graphicx,amssymb,amsmath,cite,subfigure,hyperref,calc}
\usepackage{graphicx,amssymb,amsmath,cite}

\usepackage{cite}
\usepackage[]{algorithm2e}
\theoremstyle{plain}

\theoremstyle{plain}

\theoremstyle{remark}

\newtheorem{remark}{Remark}
\usepackage{cite}
\begin{document}

\title{Inter-Layer Per-Mobile Optimization of Cloud Mobile Computing: \protect\\ A Message-Passing Approach}

\author{Shahrouz Khalili, \emph{Student Member, IEEE} and\thanks{This work was partially supported by the U.S. NSF through grant no. 1525629.}\thanks{S. Khalili and  O. Simeone are with CWCSPR, ECE Dept, NJIT, Newark, USA. E-mail: \{sk669, osvaldo.simeone\}@njit.edu.} Osvaldo Simeone, \emph{Senior Member, IEEE}}

\maketitle
\vspace{-1in}

\begin{abstract}
Cloud mobile computing enables the offloading of computation-intensive applications from a mobile device to a cloud processor via a wireless interface. In light of the strong interplay between offloading decisions at the application layer and physical-layer parameters, which determine the energy and latency associated with the mobile-cloud communication, this paper investigates the inter-layer optimization of fine-grained task offloading across both layers. In prior art, this problem was formulated, under a serial implementation of processing and communication, as a mixed integer program, entailing a complexity that is exponential in the number of tasks. In this work, instead, algorithmic solutions are proposed that leverage the structure of the call graphs of typical applications by means of message passing on the call graph, under both serial and parallel implementations of processing and communication. For call trees, the proposed solutions have a linear complexity in the number of tasks, and efficient extensions are presented for more general call graphs that include "map" and "reduce"-type tasks. Moreover, the proposed schemes are optimal for the serial implementation, and provide principled heuristics for the parallel implementation. Extensive numerical results yield insights into the impact of inter-layer optimization and on the comparison of the two implementations.
\end{abstract}

% Note that keywords are not normally used for peerreview papers.

%\IEEEpeerreviewmaketitle

\section{Introduction}
\label{sec:intro}
With the current widespread use of smart phones, there is an increasing
demand on the users' part for applications that require heavy computations
to be run on battery-powered mobile devices, such as video processing,
gaming, automatic translation, object recognition and medical monitoring. Offloading energy-consuming tasks from a mobile
device to a cloud server -- known in the literature as cyber foraging, computation offloading \cite{kumar} and, more commonly, cloud mobile
computing \cite{fern} -- provides a viable solution to this problem, as attested to by systems such as Google Voice Search, Apple Siri and
Shazam and by implementations such as MAUI \cite{maui} and ThinkAir \cite{kosta}.

\begin{figure}
\centering
\includegraphics[scale=.35]{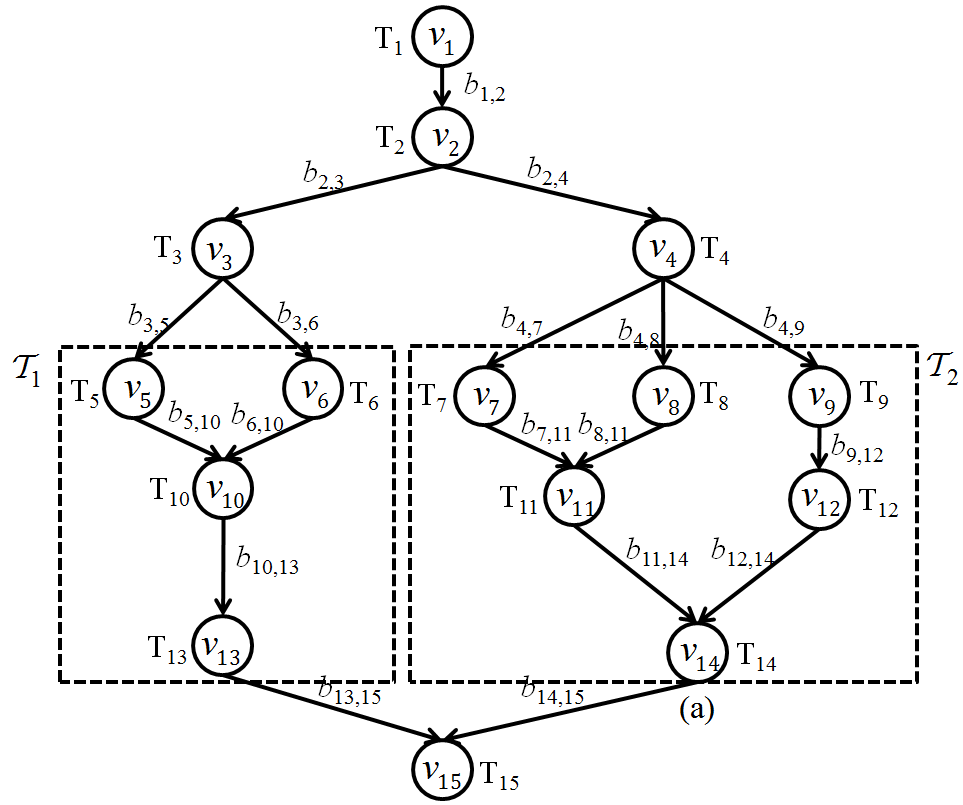}
\caption{An example of a call graph $\mathcal{G}=(\mathcal{V},\mathcal{E})$ \cite{hermp}.}\label{graph2}% including the set of data nodes $\mathcal{V}_\mathrm{D}=\{\mathrm{D}_1,...,\mathrm{D}_{|\mathcal{V}_\mathrm{D}|}\}$, denoted by squares, and the set of task nodes $\mathcal{V}=\{\mathrm{T}_1,...,\mathrm{T}_{|\mathcal{V}|}\}$, denoted by circles.}
\vspace{-1.5em}
\end{figure}

A mobile application can be partitioned into its component tasks via
profiling, producing a \emph{call graph }for the program \cite{cal}. The call graph describes the functional dependence between the different tasks (see Fig. \ref{graph2} for an example). Offloading can either take place at the coarser
granularity of entire applications, as in, e.g., \cite{sat}, or at the finer scale of individual tasks, see \cite{maui}. In the latter case, each task may be either offloaded to the cloud
or performed locally. Moreover, processing and communication processes can either be implemented one after another in a serial fashion, as assumed in most prior art, or may be parallelized in the case of non-conflicting tasks as in \cite{no}\cite{hermp}.

\textbf{State of the Art}: The large majority of prior works on the subject of optimal fine-grained
offloading tackles the problem on a per-mobile basis, and assumes a \emph{fixed physical layer}, which provides given
information rate and latency. Examples of this approach for the serial implementation include \cite{yang}, which uses a graph
partitioning formulation; \cite{odessa}, which presents a heuristic on-line approach to task
partitioning to improve latency; and \cite{sub} and \cite{cll}, which assume a time-varying channel and propose
adaptive solutions based on Lyapunov optimization and a constrained
shortest path problem, respectively. Instead, for the parallel implementation, references \cite{no}\cite{hermp} propose a dynamic programming solution, again with a fixed physical layer.
%We also refer to the review and tutorial
%papers \cite{lte,fern,gear}.

While the assumption of a fixed physical layer made in all reviewed works simplifies the problem
formulation, there is an evident interplay between decisions at the
physical layer and offloading decisions at the application layer.
Most fundamentally, the choice of the physical layer mode, e.g., of
the transmission power and information rate, determines
the mobile energy consumption, as well as the corresponding latency, for mobile-cloud communication. Therefore, a proper adaptation of the physical layer is instrumental in making cloud mobile computing viable.

Recognizing this critical interplay, more
recent work has tackled the \textit{inter-layer optimization of the physical
and of the application layers}. Specifically, references \cite{barba}\cite{stef}
 studied this problem for a general network of interfering
mobile devices by assuming \textit{coarse-grained offloading}. Fine-grained offloading is instead studied in \cite{bar}, where the authors focus on a per-mobile formulation under a serial implementation. To reduce the complexity of the resulting mixed integer program in \cite{bar}, a method is proposed that limits the exponential number of alternative offloading decisions based on feasibility arguments. Furthermore, for fixed offloading decisions, the problem is shown to have useful convexity properties. A similar problem formulation is also studied in \cite{holis}.

\textbf{Main Contributions}: In this paper, we investigate the per-mobile inter-layer fine-grained optimization of offloading decisions at the application layer and of the transmission powers at the physical layer, with the aim of minimizing energy and latency for \textit{both} serial and parallel implementations. As discussed, prior works, including \cite{bar}\cite{holis}, formulate the problem as a mixed integer program, whose complexity is exponential in the size of the call graph. Here, instead, we start from the observation that most call graphs have specific structures that can be leveraged to reduce the computational complexity. For instance, Fig. \ref{graph2} shows a typical example of an application that is composed of ``map'' tasks, which perform operations such as filtering, features extraction or sorting, and allow the successive tasks to be decomposed into independent operations (see tasks $\mathrm{T}_2$, $\mathrm{T}_3$, $\mathrm{T}_4$); along with ``reduce'' tasks, which perform summary operations such as classification or regression (see tasks $\mathrm{T}_{10}$, $\mathrm{T}_{11}$ and $\mathrm{T}_{14}$). This paper shows that, for structured graphs, solutions based on message passing can be developed for the both standard \textit{serial} implementation, (see Sec. \ref{sec:ser}), as well as the \textit{parallel} implementation (see Sec. \ref{sec:par}).

In particular, for applications with a tree structure, such as the subtrees $\mathcal{T}_1$ and $\mathcal{T}_2$ in Fig. \ref{graph2}, we develop optimal efficient message passing algorithm for the serial implementation, whose complexity is of the order $O(|\mathcal{V}|d_{in})$, where $|\mathcal{V}|$ is the number of nodes of the call graph and $d_{in}$ is the maximum in-degree. For the more challenging parallel implementation, the proposed method yields a principled suboptimal scheme whose complexity is of the same order as for the serial case. The performance of this scheme is evaluated by means of a dynamic model also introduced here. For more general call graphs, such as the one in Fig. \ref{graph2}, we generalize the proposed solutions to yield a complexity of the order $O(2^{|\mathcal{V}_s|}|\mathcal{V}|d_{in})$, where $|\mathcal{V}_s|$ is the number of nodes that, if removed, decompose the graph into subtrees (such as $\mathrm{T}_2$, $\mathrm{T}_3$ and $\mathrm{T}_4$ in Fig. \ref{graph2}, so that $|\mathcal{V}_s|=3$ for this call graph). With reference to prior work, we note that the proposed approach for parallel case generalizes the schemes in \cite{no} and \cite{hermp} by encompassing also the optimization of the physical layer. Extensive simulation results, presented in Sec. \ref{sec:simu}, bring insight into the impact of inter-layer optimization and of the call graph structure on the performance of the cloud mobile computing.

\emph{Notation}: Throughout, we use the graph terminology of, e.g., \cite{koller}. Accordingly, for a graph $\mathcal{G}=(\mathcal{V},\mathcal{E})$, a node $a$ with an incoming edge from another node $b$ is referred to as a \textit{child} of the \textit{parent} node $b$. $\mathcal{P}(n)$ and $\mathcal{C}(n)$ are the sets containing parents and children, respectively, of a node $n\in\mathcal{V}$. Given a set $\mathcal{A}\subseteq\mathbb{N} $, where $\mathbb{N} $ is the set of integers and variables $X_i$ with $i\in \mathbb{N}$, $X_{\mathcal{A}}$ is the set defined as $X_{\mathcal{A}}=\{X_i|i\in\mathcal{A}\}$; similarly, for variables $X_{i,j}$ with  $j\in\mathbb{N}$, $X_{\mathcal{A},j}$ is the set defined as $X_{\mathcal{A},j}=\{X_{i,j},i\in\mathcal{A} \}$.
\section{System Model}
\label{sys:mod}
We consider a per-mobile problem formulation in which a mobile aims at running a given application with  minimal energy expenditure and latency. For this purpose, the mobile may offload some of the computing tasks to a cloud processor, also referred to as server. We consider a configuration with a single processor both at mobile and cloud. We start in this section by introducing the key quantities at the \textit{application layer} and then at the \textit{physical layer}.
\subsection{Application Layer}
A computer application can be described by its call graph \cite{cal}. A call graph $\mathcal{G}=(\mathcal{V},\mathcal{E})$ is a \textit{directed acyclic graph}  which is used to represent the casual relation among the tasks in which a program can be partitioned. An example is shown in Fig. \ref{graph2}. Each vertex, or node, in $\mathcal{V}$ represents a particular task to be carried out within the application, e.g., data preparation, edge recognition or transform coding. We denote the task nodes as $\mathcal{V}=\{\mathrm{T}_1,...,\mathrm{T}_{|\mathcal{V}|}\}$. However, we will also use the shortcut notation $n\in\mathcal{V}$ in lieu of $\mathrm{T}_n\in\mathcal{V}$, where no confusion can arise. In the call graph $\mathcal{G}$, a directed edge $(\mathrm{T}_{m},\mathrm{T}_n)\in\mathcal{E}$ with $\mathrm{T}_{m}\in\mathcal{V}$ and $\mathrm{T}_n\in\mathcal{V}$ denotes the invocation of a ``child'' task  $\mathrm{T}_n$ by a ``parent'' task $\mathrm{T}_{m}$.

Each task node $\mathrm{T}_n$ is characterized by a parameter $v_n$, which is the number of CPU cycles required for task $\mathrm{T}_n$ to be completed. Let us define as $f^l$ and $f^r$ the number of CPU cycles/sec that can be run at the mobile (i.e., locally) and the cloud (i.e., remotely), respectively. The latency  $L^l_n= v_n/f^l$ is then the time required to compute task $\mathrm{T}_n$ locally and $L^r_n= v_n/f^r$ is the latency to run that task remotely in the case the respective processors are devoted only to the completion of task $\mathrm{T}_n$. Each edge $(\mathrm{T}_{m},\mathrm{T}_n)\in \mathcal{E}$ is instead labeled by the number of bits $b_{{m},n}$ that must be transferred by the parent task $\mathrm{T}_{m}$ in order to allow the computation of the child task $\mathrm{T}_n$.

To complete the description of the quantities of interest at the application layer, we introduce the \textit{offloading decision variables}. Specifically, we define $I_n\in\{0,1\}$ as the indicator variable that determines whether task $\mathrm{T}_n$ should be executed locally or remotely, where $I_n=0$ indicates the local execution of the task and $I_n=1$ represents the offloading of the task to the remote server. Not all the tasks may be eligible for offloading. In particular, a mobile application typically operates on input data, e.g., images or videos, that reside in the mobile device. This can be accounted for by identifying a subset $\mathcal{V}_\mathrm{D}\subseteq\mathcal{V}$ of task nodes that represent input data preparation processes, such that for every task $\mathrm{T}_m\in\mathcal{V}_\mathrm{D}$ we have $I_m=0$, i.e., local processing. These nodes are assumed to have no parents and have the role of initializing the application (see, e.g., \cite{no}\cite{hermp}). For instance, in Fig. \ref{graph2}, we may have $\mathcal{V}_\mathrm{D}=\{\mathrm{T}_1\}$. Moreover, for any graph, we assume, without loss of generality, that there is a final task to be carried out at the mobile that has no children and completes the application by, e.g., showing the results on the mobile screen. An example is task $\mathrm{T}_{15}$ in Fig. \ref{graph2} for which we then have $I_{15}=0$.

\subsection{Physical Layer}
\label{sec:intro:phy}
We now describe the parameters and the optimization variables relative to the \textit{physical layer}. The parameter $P^l$ represents the local processing power of the mobile and $P^{rf}$ is the power required to keep the mobile's RF circuits active during both transmission and reception, while $P^{rx}$ is the power needed to process the received baseband signal for decoding at the mobile. All powers are measured in Watts. The parameter $C^{dl}$ (bits/s) is the downlink capacity available to transfer the information bits from the server to the mobile. Uplink and downlink are assumed to be operated over orthogonal spectral resources.

The optimization variable $P^{ul}_{m,n}$ is the uplink power used by the mobile to transfer the necessary $b_{{m},n}$ bits in case a parent task $\mathrm{T}_{m}$ is run locally  ($I_{m}=0$) and a child task $\mathrm{T}_n$ is performed remotely ($I_n=1$) for all $(\mathrm{T}_{m},\mathrm{T}_n)\in\mathcal{E}$. Note that we allow the uplink transmit powers $P^{ul}_{m,n}$ to be different for every edge in $\mathcal{E}$, hence enabling a more flexible joint optimization of application and physical layers as in \cite{bar}. Given an uplink power $P$, we denote as
\begin{equation}
C^{ul}(P)=B\log_2 \left(1+\frac{\gamma P}{N_0B}\right)
\end{equation}
the uplink rate (bits/s) between the mobile and the server, where $\gamma$ accounts for the channel gain between mobile and the server, $B$ is the available bandwidth and $N_0$ (Watts/Hz) is noise power spectral density.

\section{Problem Formulation}
\label{sec:prob}
In this work, we aim at optimizing the application layer variables $\textbf{I}=\{I_n\}_{n=1}^{|\mathcal{V}|}$, with $I_n=0$ for $n\in\mathcal{V}_\mathrm{D}$ and for the root node, and the physical layer variables $\textbf{P}=\{P_{m,n}^{ul}\}_{(m,n)\in \mathcal{E}}$. We consider separately serial and parallel implementations.
\subsection{Serial Implementation}
\label{sec:ser}
In this section, as in most prior work, we assume that at any time, only one operation, either computation or communication, may take place, either at the mobile or at the server. Therefore, the operations needed to run a given application are performed in a serial fashion one after another. Note that the order in which these operations are scheduled is arbitrary as long as it is consistent with the procedures encoded in the call graph. For instance, for the tree $\mathcal{T}_1$ in Fig. \ref{graph2} if $I_5=I_6=I_{13}=0$ and $I_{10}=1$, tasks $\mathrm{T}_5$ and $\mathrm{T}_6$ can be first carried out in any order at the mobile; then, $b_{5,10}$ and $b_{6,10}$ bits are transferred in the uplink in any order; then, node $\mathrm{T}_{10}$ is processed at the cloud; and finally $b_{10,13}$ bits are downloaded by the mobile, which performers task $\mathrm{T}_{13}$.

Under a serial implementation, the overall latency is the sum of all the latencies required to communicate and compute across all task nodes, which can be written as (see also \cite{bar})
\begin{equation}\label{ser}
\begin{split}
L(\textbf{I},\textbf{P})&=\sum_{n=1}^{|\mathcal{V}|}L^c_{n}(I_n)+\sum_{n=1}^{|\mathcal{V}|}\sum_{m\in\mathcal{P}(n)} L^{ul}_{m,n}(I_{\{m,n\}},P^{ul}_{m,n})\\
&+\sum_{n=1}^{|\mathcal{V}|}\sum_{m\in\mathcal{P}(n)}L^{dl}_{m,n}(I_{\{m,n\}}),
\end{split}
\end{equation}
where $L^c_{n}(I_n)=(1-I_n)L^l_n+I_nL^r_n$ denotes the delay required to perform the computations associated with task $\mathrm{T}_n$ either locally or remotely; $L^{ul}_{m,n}(I_{\{m,n\} },P^{ul}_{m,n})= I_n (1-I_m) b_{m,n}/C^{ul}(P^{ul}_{m,n})$ accounts for the delay caused by the transfer of $b_{m,n}$ bits to the server if task $\mathrm{T}_n$ is offloaded ($I_n=1$) but $\mathrm{T}_m$ is not ($I_m=0$); $L^{dl}_{m,n}(I_{\{m,n\}})=(1-I_{n}) I_m  b_{m,n}/C^{dl}$ represents the latency caused by the transfer of $b_{m,n}$ bits at the mobile if $\mathrm{T}_m$ is offloaded ($I_m=1$) and $\mathrm{T}_n$ is run locally ($I_n=0$).

The energy spent by the mobile for given variables is similarly given as the sum (see also \cite{bar})
\begin{equation}\label{cost2}
\begin{split}
E(\textbf{I},\textbf{P})&=\sum_{n=1}^{|\mathcal{V}|} E^c_{n}(I_n)+\sum_{n=1}^{|\mathcal{V}|}\sum_{m\in\mathcal{P}(n)}E^{ul}_{m,n}(I_{\{m,n\}},P^{ul}_{m,n})\\
&+\sum_{n=1}^{|\mathcal{V}|}\sum_{m\in\mathcal{P}(n)}E^{dl}_{m,n}(I_{\{m,n\}}),
\end{split}
\end{equation}
where the term $E^c_{n}(I_n)=(1-I_n)P^lL^l_n$ measures the energy consumed by the mobile to perform each task $\mathrm{T}_n$ locally if $I_n=0$; the term $E^{ul}_{m,n}(I_{\{m,n\}},P^{ul}_{m,n})=(P^{ul}_{m,n}+P^{rf}) L^{ul}_{m,n}(I_{\{m,n\} },P^{ul}_{m,n})$ is the energy required, for a task $\mathrm{T}_n$ with $I_n=1$, to transfer information from all the parent tasks $m\in\mathcal{P}(n)$ that are performed locally, namely with $I_m=0$; and finally $E^{dl}_{m,n}(I_{\{m,n\}})=(P^{rf}+P^{rx}) L^{dl}_{m,n}(I_{\{m,n\}})$ is the energy consumed, for a task $\mathrm{T}_n$ with $I_n = 0$, to transfer and decode the information in the downlink from parent tasks $m\in\mathcal{P}(n)$ with $I_m = 1$.

\subsection{Parallel Operation}
\label{sec:par:op}
As an alternative to the serial operation discussed above, we now consider an implementation that allows to potentially reduce the latency by parallelizing computing and communication. This implementation was implicitly assumed in \cite{no}\cite{hermp} but without consideration for the optimization of the physical layer. According to this implementation, tasks are processed as soon as they receive the necessary information from their parents. It is then possible for uplink transmissions, downlink transmissions, local and remote computations to  occur at the same time.

As an example, consider the call tree $\mathcal{T}_2$ in Fig. \ref{graph2} with $I_7=I_8= I_9= I_{14}=0$ and $I_{11}=I_{12}=1$.
\begin{figure}
\centering
\includegraphics[scale=.4]{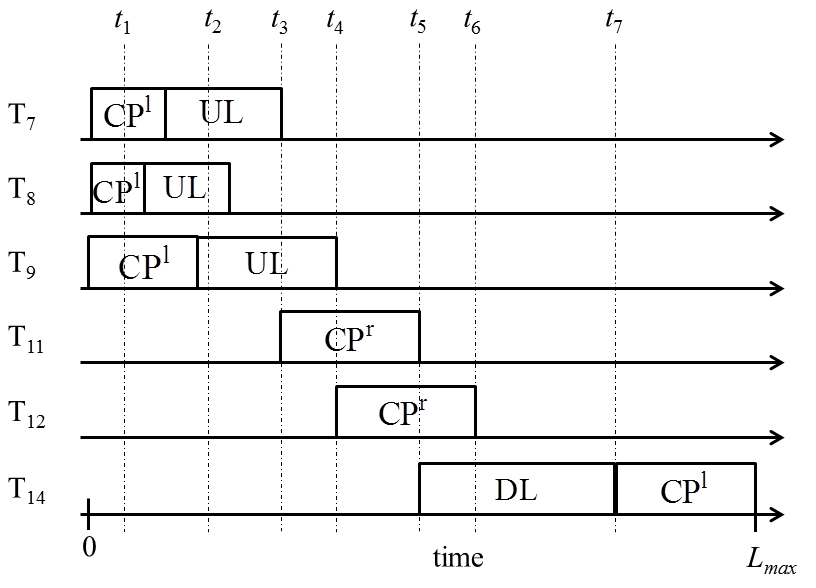}
\caption{An example of a timeline for the parallel implementation of the call tree $\mathcal{T}_2$ in Fig. \ref{graph2} with $I_7=I_8= I_9= I_{14}=0$ and $I_{11}=I_{12}=1$.}\label{time_line}
\vspace{-1.5em}
\end{figure}
An illustrative timeline is shown in Fig. \ref{time_line}, where $\mathrm{CP^l}$  denotes local computing and $\mathrm{CP^r}$ denotes remote computing; $\mathrm{UL}$ indicates that the task is uploading information bits in the uplink; and $\mathrm{DL}$ means that the task is receiving information from one or more of its parent task nodes  in the downlink. It can be seen that, for instance,  task $\mathrm{T}_{11}$ can be processed remotely as soon as the information from tasks $\mathrm{T}_7$ and $\mathrm{T}_8$ has been received by the server at time $t_3$, while uplink transmission for task $\mathrm{T}_9$ may be still ongoing. Observe that, whenever multiple concurrent uplink/downlink transfers take place at the same time, the uplink/downlink spectral resources have to be properly divided (e.g., for tasks $\mathrm{T}_7$, $\mathrm{T}_8$ and $\mathrm{T}_9$ at time $t_1$). This requires an adequate allocation of the spectral resources, such as time-frequency resource blocks in LTE. An analogous discussion applies to the computational resources.

Assuming the feasibility of allocating communication and computation resources as discussed above, the Appendix details a dynamic model that enables the evaluation of the energy and latency of the parallel implementation for given physical- and application-layer variables $\textbf{P}$ and $\textbf{I}$. This framework will be used in Sec. \ref{sec:sim} to evaluate the performance of the parallel implementation using numerical results. However, the framework in the Appendix does not lend itself  to the development of efficient optimization algorithms due to the complexity of accounting for the mentioned reallocation of the communication and computation resources. In Sec \ref{sec:par}, we develop useful heuristics for this purpose.
\subsection{Problem Formulation}
In order to optimize physician and application layer variables, we consider two different standard approaches (see, e.g, \cite{boyd}). In the first problem formulation, a weighted sum of energy and latency is minimized via the problem
\begin{equation}\label{opt2}
\begin{split}
[\mathrm{P}.1]~~\underset{\textbf{I},\textbf{P}}{\mathrm{minimize}}~E(\textbf{I},\textbf{P})+\lambda L(\textbf{I},\textbf{P}),
\end{split}
\end{equation}
where $\lambda$ is a non-negative constant that determines the trade-off between energy and latency and can be interpreted as a Lagrange multiplier. By varying $\lambda$, one can explore the trade-off between latency and energy \cite{boyd}. An alternative problem formulation is to minimize the energy (\ref{cost2}) with a latency constraint as
\begin{equation}\label{opt}
\begin{split}
[\mathrm{P}.2]~~~~~~&\underset{\textbf{I},\textbf{P}}{\mathrm{minimize}}~E(\textbf{I},\textbf{P})\\
&\textrm{subject~to}~L(\textbf{I},\textbf{P})\leq L_{max},
\end{split}
\end{equation}
where $L_{max}$ is the maximum allowed delay. Note that, in (\ref{opt2}) and (\ref{opt}), the domains of variables $\textbf{I}$ and $\textbf{P}$ are implicit. As it will be illustrated in the next sections, it is analytically convenient to tackle problem $[\mathrm{P}.1]$ for the serial implementation and problem $[\mathrm{P}.2]$ for the parallel implementation.

\begin{remark} References \cite{no}\cite{hermp} tackled problem  $[\mathrm{P}.2]$ for the parallel implementation under the assumption that the call graph is a tree or a parallel/serial combination of trees, and assuming that the physical-layer parameters $\textbf{P}$ are not subject to optimization. Moreover, the papers \cite{no}\cite{hermp} implicitly assume that parallel communication and computation do not entail a division of the available resources, hence bypassing the issue discussed above. Under these assumptions, it is shown  that the problem can be efficiently, albeit approximately, solved via dynamic programming by quantizing the set of possible delays. Reference \cite{bar} studied instead problem $[\mathrm{P}.2]$ for the serial implementation. The solution given in \cite{bar} prescribes a properly pruned exhaustive search over the variables $\textbf{I}$, and leverages the fact that, for a fixed $\textbf{I}$, the problem of optimization over $\textbf{P}$, upon a proper change of variables, is convex.
\end{remark}

\section{Optimal Task Offloading for Serial Processing}
\label{sec:ser}
In this section, we tackle problem $[\mathrm{P}.1]$ for serial processing. The key idea of the proposed approach is to leverage the factorization of the objective function in $[\mathrm{P}.1]$ in order to apply the min-sum message passing algorithm. We first detail the mentioned factorization in Sec. \ref{sec:ser:gen}. Then, in Sec. \ref{sec:ser:tree}, we discuss the proposed efficient optimal method based on min-sum message passing \cite{koller} for the special case of a call tree. Then, in Sec. \ref{sec:ser:graph}, we extend the proposed algorithm to call graphs with more general structure.
\subsection{Factorization of the Cost Function}
\label{sec:ser:gen}
The objective function for problem $[\mathrm{P}.1]$ can be factorized over the task nodes as follows:
\begin{equation}\label{p2}
\begin{split}
&\sum_{n\in\mathcal{V}}\Phi_n\left(I_{\{n\}\cup\mathcal{P}(n)},P^{ul}_{\mathcal{P}(n),n} \right),
\end{split}
\end{equation}
where the factor $\Phi_n(I_{\{n\}\cup\mathcal{P}(n)},P^{ul}_{\mathcal{P}(n),n} )$ accounts for the weighted sum of energy and latency associated with the local or the remote computation of node $\mathrm{T}_n$ and with the transmissions in uplink and/or downlink related to the edges connecting the parents of node $\mathrm{T}_n$ to node $\mathrm{T}_n$. This function is given, from (\ref{ser}) and (\ref{cost2}), as
\begin{equation}\label{phi}
\begin{split}
\Phi_n & \left(I_{\{n\}\cup\mathcal{P}(n)},P^{ul}_{\mathcal{P}(n),n} \right)=(1-I_n) P^lL^l_n+\lambda L^c_{n}(I_n)\\
&+\sum_{m\in\mathcal{P}(n)}(P^{ul}_{m,n}+P^{rf}+\lambda)L^{ul}_{m,n}(I_{\{m,n\}},P^{ul}_{m,n})\\
&+\sum_{m\in\mathcal{P}(n)}(P^{rf}+P^{rx}+\lambda)L^{dl}_{m,n}(I_{\{m,n\}}).
\end{split}
\end{equation}

We now show that the optimization in $[\mathrm{P}.1]$ over the transmission powers $\textbf{P}$ can be carried out analytically, yielding new factors that are independent of the powers. In fact, given that each power $P^{ul}_{m,n}$ appears separately in the factors of (\ref{p2}), the optimization of all powers can be carried out independently. In particular, the optimum power $\bar{P}^{ul}_{m,n}$ for all edges $(m,n)\in \mathcal{E}$ is given by the solution of the problem

\begin{equation}\label{p_ser}
\begin{split} \bar{P}^{ul}_{m,n}=\mathrm{arg~}\underset{P^{ul}_{m,n}\geq 0}{\mathrm{min}}~\frac{P^{ul}_{m,n}+P^{rf}+\lambda}{C^{ul}(P^{ul}_{m,n})}.
 \end{split}
\end{equation}
As discussed in \cite{bar}, the optimization problem in (\ref{p_ser}) becomes strictly convex with the change of variables $y_{m,n}=C^{ul}(P^{ul}_{m,n})$ and hence its unique solution can be easily found. Note that the optimum values $\bar{P}^{ul}_{m,n}$ for all $(m,n)\in\mathcal{E}$ are equal.

Substituting the optimum powers from (\ref{p_ser}) into (\ref{p2}), the problem $[\mathrm{P}.1]$ can be rewritten as
\begin{equation}\label{p22}
\begin{split}
[\mathrm{P}.1]~~&\underset{\textbf{I}}{\mathrm{minimize}}\sum_{n\in\mathcal{V}}\bar{\Phi}_n\left(I_{\{n\}\cup\mathcal{P}(n)}\right),
\end{split}
\end{equation}
where we have defined the factors
\begin{equation}
\begin{split}
&\bar{\Phi}_n\left(I_{\{n\}\cup\mathcal{P}(n)}\right)=\Phi_n\left(I_{\{n\}\cup\mathcal{P}(n)},\bar{P}^{ul}_{\mathcal{P}(n),n}\right).\label{phi_1}
\end{split}
\end{equation}
\subsection{Message Passing for a Call Tree}
\label{sec:ser:tree}
 \begin{figure}
\centering
\includegraphics[scale=.23]{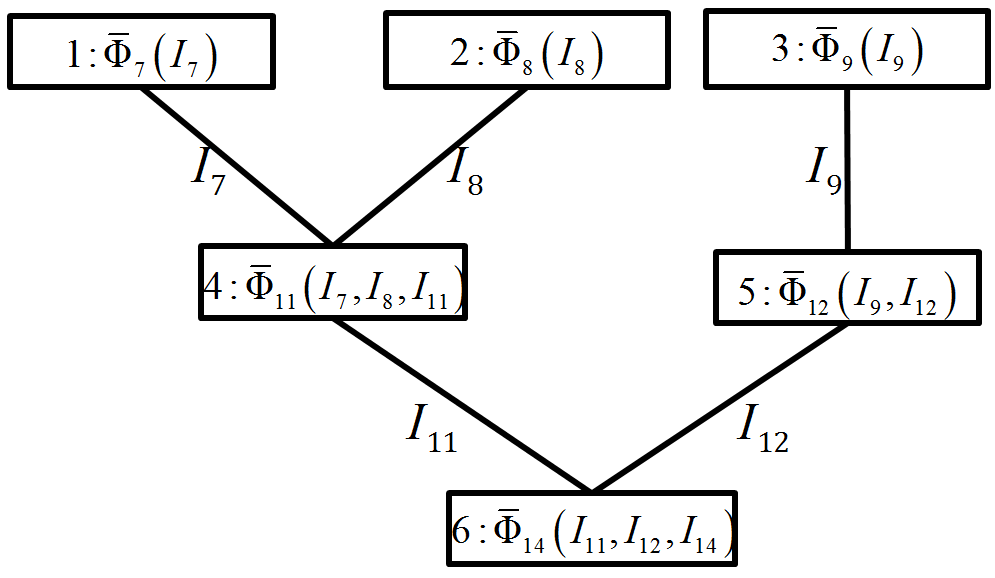}
\caption{The clique tree $\mathcal{T}_c$ corresponding to the call tree $\mathcal{T}_2$ in Fig. \ref{graph2}.}\label{cluster}
\end{figure}
For a given call tree $\mathcal{T}$, as for $\mathcal{T}_1$ and $\mathcal{T}_2$ in Fig. \ref{graph2}, the problem $[\mathrm{P}.1]$ in (\ref{p22}) can be solved exactly via the \textit{min-sum message passing} algorithm with a complexity of the order $O(|\mathcal{V}|d_{in})$, where $d_{in}$ is the maximum in-degree in the call graph. We refer to \cite{koller} for an introduction to message passing algorithms.

The algorithm operates on a clique tree $\mathcal{T}_c$ that is associated with the call tree $\mathcal{T}$. The clique tree $\mathcal{T}_c$ can be constructed from $\mathcal{T}$ as follows: (\textit{i}) replace the directed edges in $\mathcal{T}$ with undirected ones; and (\textit{ii}) substitute each task node $\mathrm{T}_n$ in $\mathcal{T}$ with a node of $\mathcal{T}_c$, which we label as the $n$th cluster node. Each cluster node $n$ is assigned the factors $\bar{\Phi}_n\left(I_{\{n\}\cup\mathcal{P}(n)}\right)$ in (\ref{phi_1}). Each edge that connects clusters $n$ and $m$ is labeled with the variable $I_m$ that appears in both clusters $n$ and $m$.  An example of a call tree and its corresponding clique tree is illustrated in Fig. \ref{cluster}.

Once the clique tree is constructed, the min-sum message passing algorithm can be directly obtained following the standard rules as detailed in \cite[Ch. 10]{koller}. To elaborate, we define $\{E^l(n), E^r(n)\}$ as the message  sent by the $n$th cluster node on the edge labeled by $I_n$, to its child cluster, where $E^l(n)$ is the value of the message corresponding to $I_n=0$ (local processing) and $E^r(n)$ is the value of the message for $I_n=1$ (remote processing). Note that the definition of the parents and children nodes follows that used for the call tree $\mathcal{T}$. The messages of the clusters that are not leaves can be calculated recursively as
\begin{equation}\label{ser_up}
\begin{split}
E^l&(n)=\sum_{m\in\mathcal{P}(n)}\mathrm{min}\left\{E^l(m)+\bar{\Phi}_n\left(I_n=0,I_m=0\right),\right.\\
&\left.E^r(m)+\bar{\Phi}_n\left(I_n=0,I_m=1\right)\right\},
\end{split}
\end{equation}
and
\begin{equation}\label{ser_up2}
\begin{split}
E^r&(n)=\sum_{m\in\mathcal{P}(n)}\mathrm{min}\Big\{E^l(m)+\bar{\Phi}_n\left(I_n=1,I_m=0\right),\\
&E^r(m)+\bar{\Phi}_n\left(I_n=1,I_m=1\right)\Big\}.
\end{split}
\end{equation}
In order to keep track of the optimal decision $\textbf{I}$, for each cluster $n$ and parent cluster $m$, we also define the functions $I^l_m(n)$ and $I^r_m(n)$, where we have $I^l_m(n)=0$ if the first argument in the min operation in (\ref{ser_up}) is smaller and $I^l_m(n)=1$ otherwise; and $I^r_m(n)$ is defined analogously with respect to (\ref{ser_up2}).
\begin{table}
 \caption{Message Passing Algorithm for the Serial Implementation}\label{tab2}
\begin{tabular}{l}
\hline
1: Calculate the powers $\bar{P}^{ul}_{m,n}$ for all\\
~~~$(m,n)\in \mathcal{E}$ using (\ref{p_ser}).\\
2: Build the corresponding clique tree as explained in Sec. \ref{sec:ser:tree} (see \\
~~~Fig. \ref{cluster}).\\
3:  \textbf{for} $n=1$:$|\mathcal{V}|$ \textbf{do}\\
~~~~~\textbf{if} $n$ is a leaf cluster\\
$~~~~~~~~E^l(n)=\begin{array}{ll}
0
\end{array}$\\
$~~~~~~~~E^r(n)=\begin{array}{ll}
\infty
\end{array}$\\
~~~~~\textbf{else}\\
~~~~~~~~~Update $E^l(n)$ and $E^r(n)$  by using (\ref{ser_up}) and (\ref{ser_up2}) and calculate\\
~~~~~~~~~$I^l_m(n)$ and $I^r_m(n)$ for all $m\in\mathcal{P}(n)$\\
~~~~~~~~~as explained in Sec. \ref{sec:ser:tree}.\\
4: Trace back the optimum decisions.\\
\hline
\end{tabular}
\end{table}

As detailed in Table \ref{tab2}, the messages are first sent by the leaf clusters, and then each cluster transmits its message $\{E^l(n),E^r(n)\}$ to its child cluster as soon as it has received the message from all its parents. The message passing algorithm is detailed in Table \ref{tab2}. The optimum decisions are finally obtained via backtracking, starting from the root node $\mathcal{V}$ so that for any node $n$ and every parent $m\in\mathcal{P}(n)$, we set $I_{m}=I_m^l(n)$ if $I_{n}=0$ and $I_m=I_m^r(n)$ otherwise. From (\ref{ser_up}) and (\ref{ser_up2}), the complexity of serial implementation is of order $O(|\mathcal{V}|d_{in})$, since every node needs to sum at most $d_{in}$ metrics, each of which only requires two sums and a binary comparison.

\subsection{Message Passing for a General Graph}
\label{sec:ser:graph}

In the case of a more general call graph $\mathcal{G}$, it is not possible to directly convert the call graph to a clique tree as done above for a call tree.

We outline here two solutions to this problem. First, assume that the call graph is such that by removing a small number subset $\mathcal{V}_\mathrm{S}$ of nodes, one can partition the graph into subtrees. This is the case for typical graphs, such as that in Fig. \ref{graph2}, with a small number of ``map'' and ``reduce'' nodes (see Sec. \ref{sec:intro}). For such graphs, similar to the observation in \cite{hermp}, one can apply message passing scheme introduced above on each subtree for all possible instantiations of the offloading decisions for the mentioned fixed nodes. Then, the minimum value of the function in (\ref{p22}) is calculated over all such instantiations. The complexity of this approach is of the order  $O(2^{|\mathcal{V}_s|}|\mathcal{V}|d_{in})$.

For graphs with an even more general structure, the junction tree algorithm can be applied to obtain a clique tree \cite[Ch. 10]{koller}. Once the clique tree is obtained, message passing can be implemented by extending the approach described in the previous subsection. The complexity of this scheme depends on the treewidth of the graph \cite{koller}. In general, unless $|\mathcal{V}_\mathrm{S}|$ is prohibitively large, the previous approach is to be preferred due to the possibility to reuse efficient algorithm in Table \ref{tab2}.

\section{Optimization of Task Offloading for Parallel Processing}
\label{sec:par}
In this section, we tackle the problem $[\mathrm{P}.2]$ in the presence of parallel processing. As for the serial case, we concentrate on call trees  in Sec. \ref{sec:par:tree}, and in Sec. \ref{sec:par:gen} we discuss the extensions to more general call graphs.

As explained in Sec. \ref{sec:prob}, in order to evaluate energy and latency of a parallel implementation, one needs to keep track of the number of concurrent processes that use the local and remote CPUs as well as the uplink and downlink bandwidth. While the dynamic model presented in the Appendix is able to do so, its use for optimization appears challenging. Hence, in this section, in order to develop a useful optimization heuristic, we assume that the number of concurrent uploads, downloads, local computations and remote computations are fixed. Under this simplifying assumption, we propose an algorithm that solves problem $[\mathrm{P}.2]$ to any arbitrary precision with linear complexity via message passing, and, specifically, via dynamic programming. The performance of the obtained heuristic solution is then evaluated by means of the dynamic model described in the Appendix.

To elaborate, we fix the number of concurrent upload and download transmissions to $N^{ul}$ and $N^{dl}$, respectively, and, the number of concurrently computed tasks locally or remotely as $N^l$ and $N^r$, respectively. The fixed values of $N^{ul}$, $N^{dl}$, $N^l$ and $N^r$ define parameters that can be set by the designer, yielding different optimization solutions that can be evaluated via the dynamic model in the Appendix. More discussion on the selection of these parameters can be found in Sec. \ref{sec:sim}.

Having fixed the mentioned parameters, the optimization proceeds as follows. To start, the available uplink and downlink capacities are obtained as
 \begin{subequations}\label{cap}
\begin{align}
&C^{ul}_{par}(P^{ul}_{m,n})=\frac{C^{ul}(N^{ul}P^{ul}_{m,n})}{N^{ul}}\\
&\mathrm{and}~C^{dl}_{par}=\frac{\log_2\left(1+(2^{C^{dl}}-1)N^{dl}\right)}{N^{dl}},
\end{align}
\end{subequations}
which correspond to the rates achievable when the spectral resources, either in the time or in the frequency, are equally divided into $N^{ul}$ and $N^{dl}$ parts, respectively. Similarly, the frequency of the local and the remote processors can be obtained by
\begin{equation}\label{far}
f^l_{par}=\frac{f^l}{N^l}~\mathrm{and}~f^r_{par}=\frac{f^r}{N^r}.
\end{equation}

Following \cite{no}, we start by observing that, for each task $\mathrm{T}_n$, the delay required to complete the tasks of the subtree in $\mathcal{G}$ rooted at any task node $\mathrm{T}_n$ can be calculated recursively, given that the completion of task $\mathrm{T}_n$ requires completion of all the parent tasks. Specifically the time $L^{(n)}_{par}(\textbf{I},\textbf{P})$ by which the subtree rooted at $\mathrm{T}_n$ is completed, given the decisions $(\textbf{I},\textbf{P})$, can be written in terms of the same quantities for its parents as
\begin{equation}\label{par}
\begin{split}
L^{(n)}_{par}(\textbf{I},\textbf{P})=&\underset{m\in\mathcal{P}(n)}{\max}\left\{L^{(m)}_{par}(\textbf{I},\textbf{P})+L^{ul}_{m,n}(I_{\{m,n\}},P^{ul}_{m,n})\right.\\
&\left.+L^{dl}_{m,n}(I_{\{m,n\}})\right\}+L^c_{n}(I_n),\\
\end{split}
\end{equation}
where the $L^{(m)}_{par}(\textbf{I},\textbf{P})$ is the latency of the subtree rooted at the parent node $\mathrm{T}_m$ and the latency terms are defined as in (\ref{ser}). % $L^{ul}_{m,n}(I_{\{m,n\}},P^{ul}_{m,n})$  and  $L^{dl}_{m,n}(I_{\{m,n\}})$ are the latencies due to the uploading or downloading of information and $L^c_{n}(I_n)$ is the latency to perform task $\mathrm{T}_n$ as they are expressed in (\ref{ser}).
Note that since $I_n=0$ for the leaf nodes in $\mathcal{V}-\mathrm{D}$, we have $L^{(n)}_{par}(\textbf{I},\textbf{P})=0$ for $n\in\mathcal{V}_\mathrm{D}$. The expression (\ref{par}) can be then calculated recursively starting from the leaf nodes, and the final delay is given by $L_{par}(\textbf{I},\textbf{P})=L^{(|\mathcal{V}|)}_{par}(\textbf{I},\textbf{P})$. %We emphasize that (\ref{par}) is an upper bound on the actual latency due to the bounds (\ref{cap})-(\ref{far}).
%\begin{equation}
%L_{par}(\textbf{I},\textbf{P})=L^{(|\mathcal{V}|)}_{par}(\textbf{I},\textbf{P}).
%\end{equation}
\subsection{Message Passing for a Call Tree}
\label{sec:par:tree}
In order to develop an approximate solution to problem $[\mathrm{P}.2]$ under the said assumptions (see (\ref{cap})-(\ref{far})), as in \cite{no}, we partition the set of possible delays into $K$ intervals by means of the quantization function
\begin{equation}\label{time_q}
\begin{split}
&q(t)=t_k~~~~\mathrm{if}~ t\in(t_{k-1},t_k],
\end{split}
\end{equation}
where $0\leq t_1\leq t_2\leq... \leq t_K=L_{max}$ are given predefined latency values. We take for simplicity $t_k=(k-1)\epsilon$ for a given quantization step $\epsilon>0$. The algorithm presented below provides  an approximation of the optimal solution of the program at hand, which, following the same arguments as in \cite{no}\cite{hermp}, become increasingly accurate as $\epsilon$ becomes smaller.

We define $\mathcal{T}_n$ as the subtree $\mathcal{G}$ that is rooted at the task $\mathrm{T}_n$. Moreover, we let $E^l(n,k)$ denote the minimum energy needed to run the the tasks in  $\mathcal{T}_n$ if node $\mathrm{T}_n$ is executed locally and under the constraint that the latency is less than $t_k$. Note that the energy $E^l(n,k)$ is minimized with respect to the offloading variables in vector $\textbf{I}$ corresponding to the task nodes in the mentioned subtree except $\mathrm{T}_n$, as well as over the uplink powers in vector $\textbf{P}$ corresponding to all the edges within the subtree. Similarly, we define  $E^r(n,k)$ as the minimum energy cost for $\mathcal{T}_n$ if $\mathrm{T}_n$ is performed remotely and under the delay constraint $t_k$. We also correspondingly define the set $\mathcal{I}^l(n,k)=\{I^l_m(n,k)\}_{m\in\mathcal{P}(n)}$ that  contains the optimum offloading decisions for the parent nodes $\mathrm{T}_m$  of node $\mathrm{T}_n$ if the latter is performed locally under the latency $t_k$ for the subtree rooted at $\mathrm{T}_n$. Similarly, we define $\mathcal{I}^r(n,k)=\{I^r_m(n,k)\}_{m\in\mathcal{P}(n)}$ as the set containing the optimum decisions for the parent nodes $\mathrm{T}_m$ of node $\mathrm{T}_n$, if the latter is performed remotely with the latency constraint $t_k$.

The proposed dynamic programming algorithm computes the cost functions $E^l(n,k)$ and $E^r(n,k)$  and the sets $\mathcal{I}^l(n,k)$ and $\mathcal{I}^r(n,k)$ recursively from the energy cost functions $E^l(m,j)$ and $E^r(m,j)$ of all the parent nodes $m\in\mathcal{P}(n)$ under all the delay constraints $t_j$ with $j=1,...,k-1$. Specifically, we set $E^l(n,k)=\infty$ and  $E^r(n,k)=\infty$ for  $k\leq0$. We can then obtain the recursive relationship
\begin{equation}\label{update}
\begin{split}
E^l&(n,k)=\\
&P^lL^l_n+\sum_{m\in\mathcal{P}(n)}\mathrm{min}\left\{E^l\Big(m,k-Q(L^l_n)\Big),\right.\\
&\left.E^r\left(m,k-Q\left(L^l_n+\frac{b_{m,n}}{C^{dl}_{par}}\right)\right)+(P^{rf}+P^{rx})\frac{b_{m,n}}{C^{dl}_{par}}\right\},
\end{split}
\end{equation}
where the function $Q$ is defined as $Q(t)=k~\mathrm{if}~ t\in[t_{k-1},t_{k})~\mathrm{~for~all}~ k\in\{1,...,K\}$.

Equation (\ref{update}) accounts for the fact that the minimum energy cost required to run the task in the subtree $\mathcal{T}_n$ within a latency $t_k$ if $\mathrm{T}_n$ is run locally is given by the sum of the local processing energy $P^lL^l_n$ (see $E^c_n(I_n)$ in (\ref{cost2})) and of the energies required to run all the subtrees $\mathcal{T}_m$ with $m\in\mathcal{P}(n)$. For the latter, each parent node $\mathrm{T}_m$ can be run either locally, requiring energy  $E^l(m,k-Q(L^l_n))$, or remotely, with an energy $E^r(m,k-Q(L^l_n+\frac{b_{m,n}}{C^{dl}_{par}}))$. We observe that, if node $\mathrm{T}_m$ is performed locally, the latency allowed for the subtree $\mathcal{T}_m$ is $t_k-q(L^l_n)$ and hence the corresponding minimum energy is $E^l(m,k-Q(L^l_n))$, and similarly for the case in which $\mathcal{T}_m$ is carried out remotely the energy can be calculated as in (\ref{update}). In (\ref{update}), the $\mathrm{min}\{\cdot,\cdot\}$ operation accounts for the choice of whether node $\mathrm{T}_n$ should be performed locally or remotely. Accordingly, the set $\mathcal{I}^l(n,k)=\{I^l_m(n,k)\}_{m\in\mathcal{P}(n)}$ can be evaluated during calculation of $E^l(n,k)$ in (\ref{update}) by observing which term in the function $\mathrm{min\{\cdot,\cdot\}}$  is smaller. Specifically, we can write $I^l_m(n,k)=0$ if the first term is smaller and $I^l_m(n,k)=1$ otherwise.

Similar to (\ref{update}), we can also write
\begin{equation}\label{opt_c}
\begin{split}
%E^l&(n,t_k)=\infty~~\mathrm{if} ~t_k-q(L^l_n)<0\\
E^r&(n,k)=\sum_{m\in\mathcal{P}(n)}\mathrm{min}\Bigg\{\Bigg((\bar{P}^{ul}_{m,n,k}+P^{rf})\frac{b_{m,n}}{C^{ul}_{par}(\bar{P}^{ul}_{m,n,k})}\\
&+E^l\left(m,k-Q\left(L^r_n+\frac{b_{m,n}}{C^{ul}_{par}(\bar{P}^{ul}_{m,n,k})}\right)\right)\Bigg),\\
&E^r\Big(m,k-Q(L^r_n)\Big)\Bigg\},\\
\end{split}
\end{equation}
where uplink $\bar{P}^{ul}_{m,n,k}$ is selected as detailed below. The two arguments of the $\mathrm{min\{\cdot,\cdot\}}$ operator measures the energy cost of the subtree $\mathcal{T}_m$ in the case that the parent node $\mathrm{T}_m$ is performed locally or remotely, respectively, and are explained in an analogous fashion as for (\ref{update}). Furthermore, the set $\mathcal{I}^r(n,k)=\{I^r_m(n,k)\}_{m\in\mathcal{P}(n)}$ can be evaluated during calculation of $E^r(n,k)$ in analogous fashion as $I^l_m(n,k)$.

Once equations (\ref{update})-(\ref{opt_c}) are evaluated starting from the leaf nodes  of $\mathcal{G}$ to the root, the optimum powers $\textbf{P}$ and offloading decisions $\textbf{I}$ are obtained via backtracking from the root to the leaves of $\mathcal{G}$. Specifically, since the root node must be performed locally  within the delay constraint $L_{max}$, the optimum solution ($\textbf{I}$,$\textbf{P}$) can be found starting from the optimal decisions associated with $E^l(|\mathcal{V}|,L_{max})$ by keeping track  of  the maximum allowed delay $t_n$ for each subtree $\mathcal{T}_n$. The complete dynamic complete programming algorithm is presented in Table \ref{al} and the backtracking method is explained in Table \ref{trace}.
\begin{table}
 \caption{Dynamic Programming Solution for Parallel Implementation}\label{al}
\begin{tabular}{l}
\hline
1:  \textbf{for} $n=1$:$|\mathcal{V}|$ \textbf{do}\\
~~~~~\textbf{if} $\mathrm{T}_n\in \mathcal{V}_\mathrm{D}$\\
$~~~~~~~E^l(n,k)=\begin{array}{ll}
0 & \mathrm{for~all~} k \\
\end{array}$\\
$~~~~~~~E^r(n,k)=\begin{array}{ll}
\infty  & \mathrm{for~all~} k \\
\end{array}$\\
~~~~~\textbf{else}\\
~~~~~~~\textbf{for} $k=1$, $K$ \textbf{do}\\
~~~~~~~~~Calculate the powers $\bar{P}^{ul}_{m,n,k}$ for all $(m,n)\in \mathcal{E}$ using (\ref{p_u2}).\\
~~~~~~~~~Update $E^l(n,k)$, $E^r(n,k)$, $\mathcal{I}^l(n,k)$ and $\mathcal{I}^r(n,k)$ by using\\
~~~~~~~~~~(\ref{update})-(\ref{opt_c}).  \\
2: Trace back the optimum decisions from $E^l(|\mathcal{V}|,k)$ using the\\
~~~algorithm in Table \ref{trace}.\\
\hline
\end{tabular}
\end{table}

Optimization of the powers is carried out by observing that, thanks to the decomposition made possible by dynamic programming, the powers  $P^{ul}_{m,n,k}$ appear in separate terms in (\ref{opt_c}). Therefore, without loss of optimality, the powers $P^{ul}_{m,n,k}$ can be optimized separately from each term in (\ref{opt_c}). This optimization is complicated by the presence of the non-differentiable term  $Q(L^r_n+\frac{b_{m,n}}{C^{ul}_{par}(\bar{P}^{ul}_{m,n,k})})$. To address this issue, for each $(m,n)\in\mathcal{E}$ and each $k\in\{1,...,K\}$ we calculate
\begin{equation}\label{p_u2}
\bar{P}^{ul}_{m,n,k}=\mathrm{arg}~\underset{P^{ul}_{m,n}\geq 0}{\mathrm{min}}~E^r(n,k,P^{ul}_{m,n}),
\end{equation}
where
\begin{equation}\label{p_u}
\begin{split}
E^r(n,k,P^{ul}_{m,n})&\triangleq(P^{ul}_{m,n}+P^{rf})\frac{b_{m,n}}{C^{ul}_{par}(P^{ul}_{m,n})}\\
&+E^l\left(m,k-Q\left(L^r_n+\frac{b_{m,n}}{C^{ul}_{par}(P^{ul}_{m,n})}\right)\right).
\end{split}
\end{equation}
by solving $k-Q(L^r_n)+1$ convex subproblems. To this end, we note that the equality $Q(L^r_n+b_{m,n}/C^{ul}_{par}(P^{ul}_{m,n}))=j$ holds as long as the inclusion $ P^{ul}_{m,n}\in \mathcal{R}_{m,n,j}$ is satisfied with
 \begin{equation}\label{r}
\begin{split}
&\mathcal{R}_{m,n,j}= \left(\left(2^{\frac{b_{m,n}}{B(t_{j}-L^r_n)}}-1\right)/\gamma',\left(2^{\frac{b_{m,n}}{B(t_{j-1}-L^r_n)}}-1\right)/\gamma'\right],
%&~~~~~~~~~~~~~~~~~~~~~~~~~~~~~~~~~~\mathrm{for~ all~} j\in \{1,2,...,K\}|t_j>L^r_n \mathrm{~and~}t_{j-1}>L^r_n.
\end{split}
\end{equation}
where we defined $\gamma'=\frac{\gamma N^{ul}}{BN_0}$. We can then calculate $\bar{P}^{ul}_{m,n,k}$ in (\ref{p_u2}) by first solving the problems
\begin{equation}\label{opt_p}
P^{ul}_{m,n,j}=\mathrm{arg}~\underset{P^{ul}_{m,n}\in \mathcal{R}_{m,n,j}}{\mathrm{min}}~(P^{ul}_{m,n}+P^{rf})\frac{b_{m,n}}{C^{ul}_{par}(P^{ul}_{m,n})},
\end{equation}
for all $j\in\{Q(L^r_n),...,k\}$ and then set
\begin{equation}\label{opt_p2}
\begin{split}
\bar{P}^{ul}_{m,n,k}&=\mathrm{arg~}\underset{j\in\{Q(L^r_n),...,k\} }{\mathrm{min}}~(P^{ul}_{m,n,j}+P^{rf})\frac{b_{m,n}}{C^{ul}_{par}(P^{ul}_{m,n,j})}\\
&+E^l\left(m,k-Q\left(L^r_n+\frac{b_{m,n}}{C^{ul}_{par}(P^{ul}_{m,n,j})}\right)\right).
\end{split}
\end{equation}
Each problem (\ref{opt_p}) becomes convex by means of the change of variable $y_{m,n}=C^{ul}_{par}(P^{ul}_{m,n})$ \cite{bar}.

Since the maximum number of convex optimizations that need to be solved at each time instant for each node can be upper bounded by $d_{in}K$, and $K$ is proportional to $1/\epsilon$, the complexity of the proposed algorithm in Table \ref{al} is given by $O(|\mathcal{V}|d_{in}/\epsilon^2)$.
\begin{table}
 \caption{Backtracking algorithm for Table \ref{al}}\label{trace}
\begin{tabular}{l}
\hline
1: Set $L_{|\mathcal{V}|}=L_{max}$ and $I_{|\mathcal{V}|}=0.$\\
2:~~~\textbf{for} $n=|\mathcal{V}|:1$ \textbf{do} \\
~~~~~~~~~~\textbf{for} all $m\in\mathcal{P}(n)$ \textbf{do} \\
~~~~~~~~~~~~~\textbf{if} $I_{n}=0$\\
~~~~~~~~~~~~~~~~~~\textbf{if} $I^l_m(n,Q(L_n))=0$\\
~~~~~~~~~~~~~~~~~~~~~~Set $I_m=0$ and  $L_m=L_n-L^l_n$.\\
~~~~~~~~~~~~~~~~~~\textbf{else}\\
~~~~~~~~~~~~~~~~~~~~~~Set $I_m=1$ and  $L_m=L_n-\left(L^l_n+\frac{b_{m,n}}{C^{dl}_{par}}\right)$.\\
%~~~~~~~~~~~~~~~~~~\textbf{end if} \\
~~~~~~~~~~~~~\textbf{else}\\
~~~~~~~~~~~~~~~~~~\textbf{if} $I^r_m(n,Q(L_n))=0$\\
~~~~~~~~~~~~~~~~~~~~~~Set $I_m=0$, $\bar{P}^{ul}_{m,n}=\bar{P}^{ul}_{m,n,Q(L_n)}$\\
~~~~~~~~~~~~~~~~~~~~~~and  $L_m=L_n-\left(L^r_n+\frac{b_{m,n}}{C^{ul}_{par}(\bar{P}^{ul}_{m,n})}\right)$ .\\
~~~~~~~~~~~~~~~~~~\textbf{else}\\
~~~~~~~~~~~~~~~~~~~~~~Set $I_m=1$ and  $L_m=L_n-L^r_n$.\\
%~~~~~~~~~~~~~~~~~~\textbf{end if} \\
%~~~~~~~~~~~~~\textbf{end if}\\
%~~~~~~~~~~\textbf{end for}\\
%~~~~~\textbf{end for}\\
\hline
\end{tabular}
\end{table}
\subsection{Message Passing for a General Call Graph}
\label{sec:par:gen}
Similar to Sec. \ref{sec:ser:graph}, for a graph with the structure discussed in Sec. \ref{sec:intro}, the problem $[\mathrm{P}.2]$ can be solved, for fixed parameters $N^l$, $N^r$, $N^{ul}$ and $N^{dl}$, by means of an exhaustive search over the offloading decisions of the nodes that, when removed, decompose the graph into disjoint trees. Following the discussion in Sec. \ref{sec:ser:graph}, the resulting solution has a complexity of order $O(2^{|\mathcal{V}_s|}|\mathcal{V}|d_{in}/\epsilon^2)$.

\section{Simulation Results}
\label{sec:simu}
\label{sec:sim}
\begin{figure}
\centering
\includegraphics[scale=.3]{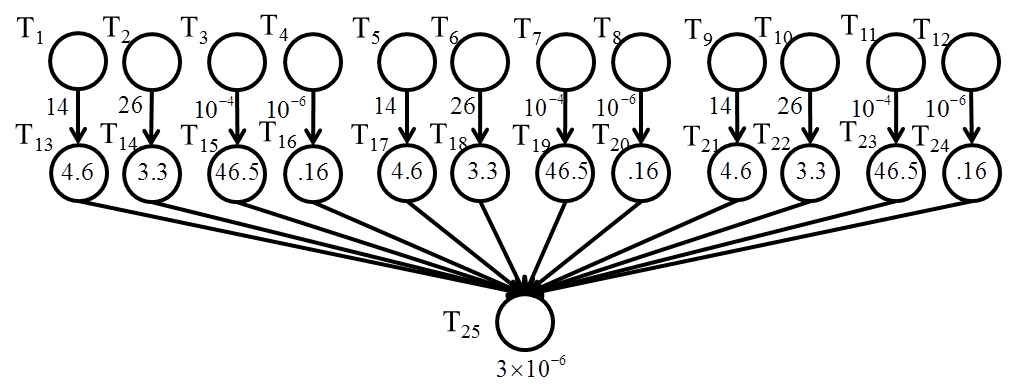}
\caption{The call tree graph used for the examples in Fig. \ref{res1}-\ref{time}. The numbers shown next to the edges that are connected to the input task nodes represent the sizes of input bits $b_{m,n}$ in Mbits and the numbers in the task nodes (circles) represent the number of CPU cycles $v_n$ normalized by $10^9$ CPU cycles (empty circles with $v_1=...=v_{12}=0$). The remaining values for case (a) are: $b_{13,25}=7.3\times10^9$, $b_{14,25}=1.4\times10^3$, $b_{15,25}=1.4\times10^3$, $b_{16,25}=1.4\times10^7$ bits,  $b_{17,13}=b_{21,25}=b_{13,25}$, $b_{18,25}=b_{22,25}=b_{14,25}$, $b_{19,25}=b_{23,25}=b_{15,13}$ and $b_{20,25}=b_{24,25}=b_{16,25}$. In case (b), all the parameters are the same as case (a) except for $b_{3,15}=b_{4,16}=b_{7,19}=b_{8,20}=b_{11,23}=b_{12,24}=11.4$ Mbits, $b_{14,25}=b_{15,25}=b_{16,25}=b_{18,25}=b_{19,25}=b_{20,25}=b_{22,25}=b_{23,25}=b_{24,25}=14.6\times10^7$ bits, $b_{13,25}=b_{17,25}=b_{21,25}=7.3\times10^7$ bits and $v_{15}=v_{19}=v_{23}=4.6\times10^9$, $v_{16}=v_{20}=v_{24}=3.6\times10^9$ and $v_{25}=3.42\times10^9$ CPU cycles.}\label{g_ex}
\vspace{-1.5em}
\end{figure}

In this section, we provide some numerical example based on the analysis developed in the previous sections. We start by considering the call tree in Fig. \ref{g_ex} in order to simplify the interpretation of the results and gain an insight into the performance of the considered techniques. In this example, $\mathrm{T}_{13},...,\mathrm{T}_{24}$ process input data present at the mobile device, represented by nodes $\mathcal{V}_\mathrm{D}=\{\mathrm{T}_1,...,\mathrm{T}_{12}\}$, e.g., to extract some features, and then root node $\mathrm{T}_{25}$ performs a ``reduce'' operation, such as classification, on the extracted features at the mobile ($I_{25}=0$). We set  $P^l=0.4$ Watts, which is a common for smart phones \cite{no,cpu_url,cpu_url2}; $f^l=10^9$ CPU cycles/s (e.g., Apple iPhone 6 processor has maximum clock rate of 1.4 Ghz); $f^r=10^{10}$ CPU cycles/s (e.g., AMD FX-9590 has a clock rate of 5 Ghz \cite{amd}); $\gamma/(BN_0)=27$ dB, $P^{rf}=0$ W, $P^{rx}=0$ W, $B=1$ MHz,  $C^{dl}=200$ Mbits/s unless stated otherwise. For both the serial implementation (solid lines) and the parallel implementation (dashed lines), optimization is performed according to the algorithms described in Sec. \ref{sec:ser} and Sec. \ref{sec:par}, respectively, and, for the parallel implementation, the performance is evaluated using the dynamic model presented in the Appendix with step size $\epsilon_d=0.1$. For parallel optimization, we set $N^{ul}=N^{dl}=N^l=N^r$ in (\ref{cap}) and (\ref{far}) to an optimized value in the range $[1,4]$ and we have $\epsilon=0.1$. Note that the performance of the optimization was found not to be significantly improved with smaller values of $\epsilon$ and not to be increased by choosing larger values for $N^{ul}=N^{dl}=N^l=N^r$.

\begin{figure}
\centering
\includegraphics[scale=.33]{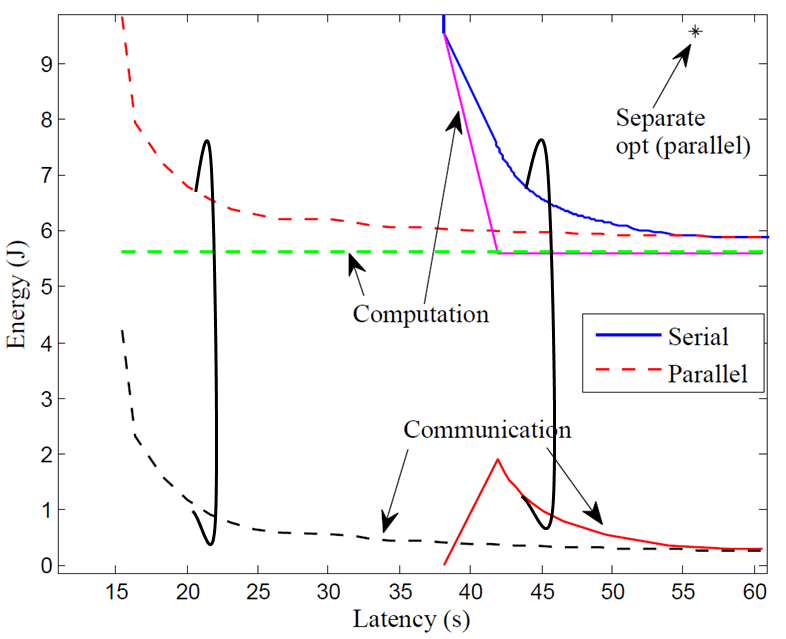}
\caption{Energy and latency trade-off for the call graph $\mathcal{G}$ in Fig. \ref{g_ex} (case (a)). The program can be completely performed locally with $E=65.6$ J and $L=164$ s. Moreover, separate optimization for serial implementation yields  $E=9.7$ J and $L=178$ s.}\label{res1}
\end{figure}

In Fig. \ref{res1}, the mobile energy cost for the serial and the parallel implementations are plotted versus the latency, along with their communication and computation components for the graph in Fig. \ref{g_ex} with the selection of parameters marked as case (a) in the caption of Fig. \ref{g_ex}. The parameters of the graph are chosen to yield the same range of latencies and energy consumptions as in \cite{maui} and \cite{hermp}. With the selected parameters, performing the application locally requires an energy equal to $65.6$ J and has a latency of $164$ s (outside the range of Fig. \ref{res1}). Fig. \ref{res1} shows that significantly smaller latencies and energy expenditures can be obtained by properly optimizing the offloading decisions and the communication strategy. For instance, with an energy expenditure of $6.5$ J, an optimized parallel implementation yields a latency of around $20$ s, while an optimized serial implementation requires a latency of around $45$ s.

The parallel implementation is shown here to have the potential to strictly outperform the serial implementation and to enable the operation at latencies that are unattainable with the serial implementation. Moreover, as the latency increases, the energy can be seen to decrease mostly due to the fact that the communication powers can be reduced. An exception to this trend is observed for the serial implementation around the latency $L=42$ s, due to the fact that the optimum application layer decisions prescribe more tasks to be offloaded for $L\geq 42$ s.

In order to provide a further reference performance for inter-layer optimization, we consider a conventional \textit{separate design} strategy, whereby: (\textit{i}) the uplink transmission power for each task is obtained by imposing the constraint that transmitting in the uplink require a time no larger than that necessary to perform that task locally (see \cite[Sec. 3]{bar} for a similar approach); (\textit{ii}) the optimization of the offloading decisions is carried out by following the proposed algorithms with a fixed physical layer, which amount to the schemes in \cite{no}\cite{hermp} for the parallel implementations. For the serial implementation, this separate approach yields a latency of $178$ s and an energy expenditure of $9.7$ J, which is outside the range of Fig. \ref{res1}, while for parallel processing the observed energy-latency power is illustrated in this figure. Note that separate optimization does not attempt to adapt the physical layer to the application layer requirements and hence it yields a single energy-latency point in the considered latency range.

Fig. \ref{no-gain} shows the energy-latency trade-off for the call graph in Fig. \ref{g_ex} for both case (a) and case (b) as detailed in the caption of Fig. \ref{g_ex}. Note that the separate optimization for case (b) with the parallel implementation yields $E=22.5$ J for $L=38.5$, which is out of the range of Fig. \ref{no-gain}. The results in Fig. \ref{no-gain} suggest that the gains offered by the parallel implementation over the serial implementation depend strongly on the chosen call graph.

To gain more insight into this point, Fig. \ref{time} illustrates the timeline corresponding to the parallel implementation for case (a) and case (b) for $L=20$ s. Here, we use the same definition for $\{\mathrm{ID},\mathrm{CP^l},\mathrm{CP^r},\mathrm{UL},\mathrm{DL}\}$ as in Fig. \ref{time}. It can be seen that in case (a), several communication and computation operations take place in parallel for a significant  fraction of the time, and hence the parallel implementation is advantageous as compared to the serial implementation. Instead, for case (b) most of the time is spent for uplink transmissions and hence the opportunities for parallel processing are much reduced.

In order to complement the insight obtained from the study of the call graph in Fig. \ref{res1}, here we elaborate on the impact of the structure of the call graph by considering the graph in Fig. \ref{graph2}. We plot the performance of the serial and parallel implementations for the call graph $\mathcal{G}$ as well as for the subtrees $\mathcal{T}_1$ and $\mathcal{T}_2$ in Fig. \ref{dep}. The relative values of the parameters in the call graph $\mathcal{G}$ is obtained from \cite{hermp}, and their exact values are defined in the caption of this figure. As expected, the energy required to run the application for a given latency increases as one considers a larger call graph. More importantly, the opportunities for concurrent computations and communications are enhanced on larger subgraphs, and, as a result, for $\mathcal{T}_2$ and $\mathcal{G}$, parallel processing provides more substantial gain over the serial implementation than in $\mathcal{T}_1$.

\section{Concluding Remarks}

In this paper, we studied the inter-layer optimization of cloud mobile computing systems over the power  allocation at the physical layer and offloading decisions at the application layer with the aim of exploring the achievable trade-offs between the mobile energy expenditure and latency. Unlike prior work in which the problem is formulated as a mixed integer program, here we proposed a message-passing framework that leverage the typical structure of call graphs to drastically reduce complexity. In particular, we focused on call graphs that can be decomposed into combination of a small number of subtrees when fixing the decisions of a subset of nodes, obtaining a complexity that grows exponentially only in the size of such set of nodes rather the size of the call graph. Moreover, unlike prior art, the framework is applied to both the conventional serial implementation and a parallel implementation that enables the concurrent schedule of communication and computation. Via simulation results, we demonstrated the impact of the call graph structure on the relative performance of the parallel and serial implementations, and shed light on the impact of inter-layer optimization.
\label{sec:con}

\section{Acknowledgements}

The authors would like to thank Gesualdo Scutari from University of
Buffalo for interesting discussions.

%\begin{appendices}
\appendix
\setcounter{secnumdepth}{-1}
\section{Evaluating Energy and Latency For the Parallel Implementation}
\setcounter{secnumdepth}{-1}
\label{app:eva}
In Sec. \ref{sec:par}, we proposed an analytically convenient approximation for the energy and latency of the parallel implementation. Here, we develop a dynamic model that enables the evaluation of upper bounds on the energy and latency of the parallel implementation for a fixed set of variables $(\textbf{I}$,$\textbf{P})$ by tracking the state of each task over time. To this end, we quantize the time axis similar to (\ref{time_q}) with a generally different time step $\epsilon_d$. By construction, the upper bounds calculated here become increasingly tighter as the quantization step $\epsilon_d$ decreases.

\begin{figure}
\centering
\includegraphics[scale=.27]{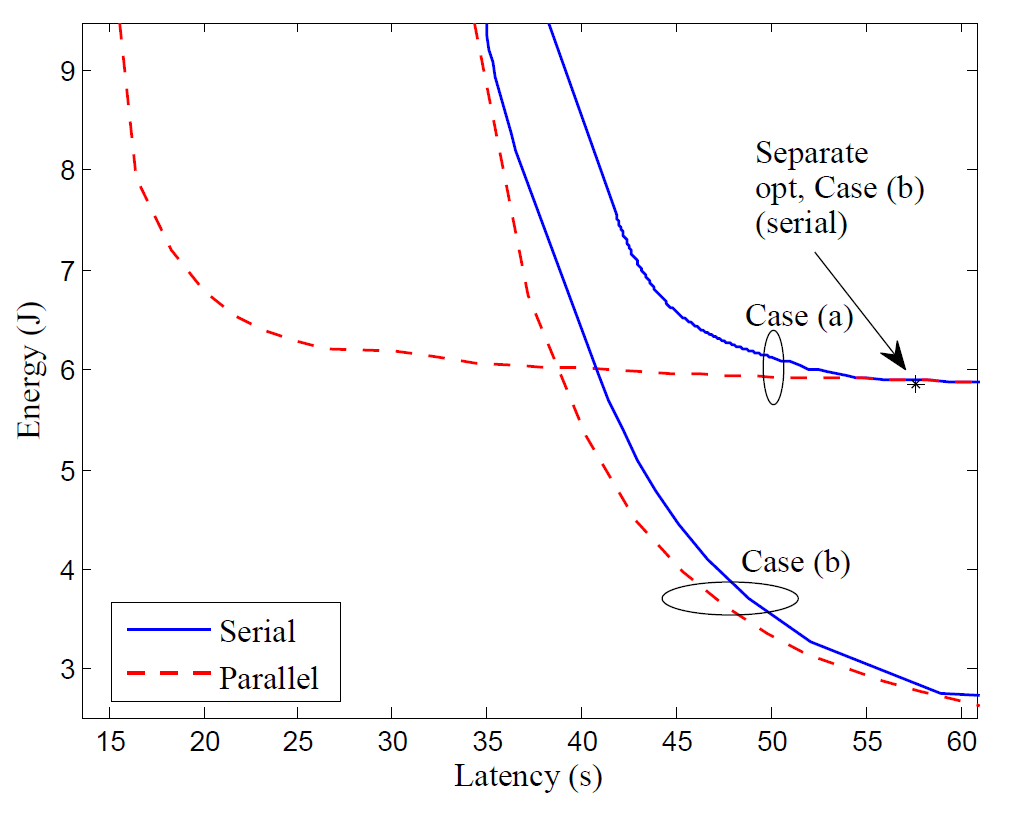}
\caption{Energy and latency trade-off for the call graph $\mathcal{G}$ in Fig. \ref{g_ex} for case (a) and case (b). Separate optimization for the parallel implementation yields $E=22.5$ J and  $L=38.5$ s for case (b) (not shown).}\label{no-gain}
\end{figure}
Define as $X_n(k)$ the state of task node $\mathrm{T}_n$ at time instant $t_k=(k-1)\epsilon_d$. The state of each node remains constant in the time range $(t_k,t_{k+1}]$ and may take any value in the set $\{\mathrm{ID},\mathrm{CM},\mathrm{CP^l},\mathrm{CP^r},\mathrm{UL},\mathrm{DL}\}$, where $\mathrm{ID}$ indicates that a task is idle in the sense that it has not started processing yet. Instead, $\mathrm{CM}$ indicates that a task is completed in terms of processing and uplink/downlink communication and other state are defined in Sec. \ref{sec:par:op}. For all $n\in\mathcal{V}_\mathrm{D}$, we initialize the state as $X_n(1)=\mathrm{CP^l}$.

To keep track of the state of the uplink and downlink transmissions, we define the following variables. The variable $b^{ul}_n(k)$ indicates the remaining information bits that task $\mathrm{T}_n$ still needs to send in the uplink at time $t_k$. For $k=1$, we have $b^{ul}_n(1)=b_{n,\mathcal{C}(n)}$ for all tasks $\mathrm{T}_n$ that are not directly connected to a leaf node with $I_n=0$ and $I_{\mathcal{C}(n)}=1$; instead, if $I_n=1$ and $\mathcal{P}(n)\in\mathcal{V}_\mathrm{D}$, we set $b^{ul}_n(k)=b_{\mathcal{P}(n),n}$; and we have $b^{ul}_n(k)=0$ otherwise. Similarly, the variable $b^{dl}_{m,n}(k)$ for $m\in\mathcal{P}(n)$ represents the remaining output bits of task $\mathrm{T}_m$ that task $\mathrm{T}_n$ needs to receive in the downlink at time $t_k$. For $k=1$, we have $b^{dl}_{m,n}(1)=b_{m,n}$ for all pairs $(m,n)$ such that $I_n=0$ and $I_{m}=1$, and $b^{dl}_{m,n}(1)=0$ otherwise.

In order to track the state of the tasks in terms of computations, we define as $c^l_n(k)$ the number of CPU cycles that are left at time $t_k$ to finish a task $\mathrm{T}_n$ with $I_n=0$, while $c^r_n(k)$ denotes the corresponding number of remaining CPU cycles for a task $\mathrm{T}_n$ with $I_n=1$. Thus, we have $c^l_n(1)=v_n$  if $I_n=0$ and $c^r_n(1)=v_n$ if $I_n=1$, while we set $c^l_n(1)=c^r_n(1)=0$ otherwise.

Let us define $N^l(k)$ as the number of tasks that are running locally and  $N^r(k)$ as the number of tasks that are running remotely at time $t_k$. Similarly, we define  $N^{ul}(k)$ and  $N^{dl}(k)$ as the number of concurrent uplink  and downlink transmissions at time $t_k$, respectively. In the proposed approach, as described below, we update the state $X_n(k)$ of each task node by making the assumption that the quantities $N^l(k)$, $N^r(k)$, $N^{ul}(k)$ and $N^{dl}(k)$ remain constant through the time interval  $(t_k, t_{k+1}]$. As argued below, this lead to the desired upper bounds on energy and latency. In the following, we treat separately the state update of each task $\mathrm{T}_n$ in any interval $(t_k,t_{k+1}]$ depending on the state $X_n(k)$ at time $t_k$.

If $X_n(k)=\mathrm{UL}$,  the amount of information that can be transmitted to the server in the time slot  $(t_k, t_{k+1}]$ should be calculated in order to update the variable $b_n^{ul}(k)$. If $I_n=1$ we have  $b^{ul}_n(k+1)=[b^{ul}_n(k)-(C^{ul}(N^{ul}(k)\bar{P}_{\mathcal{P}(n),n})/N^{ul}(k))\epsilon]^+ $ due to the uploading of information from the connected leaf node, where $[x]^+$ is equal to $x$ if $x>0$ and $x$ is equal to $0$ otherwise. Instead, if $I_n=0$, we have  $b^{ul}_n(k+1)=[b^{ul}_n(k)-(C^{ul}(N^{ul}(k)\bar{P}_{n,\mathcal{C}(n)})/N^{ul}(k))\epsilon]^+$,  due to the uploading of information to the  child task $\mathrm{T}_{\mathcal{C}(n)}$. As a result, the state of the node changes as
\begin{equation}\label{stat1}
X_n(k+1)=\left\{\begin{array}{ll}
\mathrm{UL} & \mathrm{\textbf{if}~} b^{ul}_n(k+1)>0\\
\mathrm{CM} & \mathrm{\textbf{if}~} I_n=0 ~\mathrm{and} ~b^{ul}_n(k+1)=0\\
\mathrm{CP^r} &\mathrm{\textbf{if}~} I_n=1 ~\mathrm{and} ~b^{ul}_n(k+1)=0\\
\end{array}\right.,
\end{equation}
since when $I_n=0$, the task is completed, and when $I_n=1$, the task $\mathrm{T}_n$ needs to be computed remotely.

\begin{figure}
\centering
\begin{minipage}{.8\linewidth}
\includegraphics[width=6.38cm,height=4.95cm]{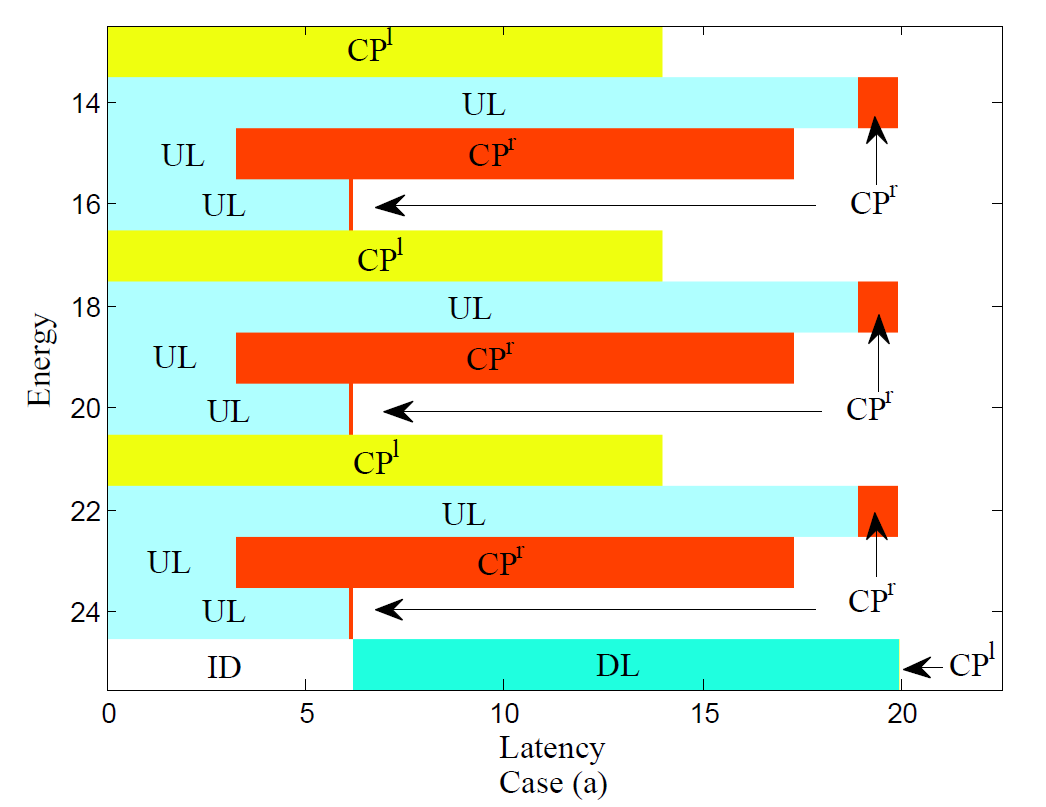}
\end{minipage}
\begin{minipage}{.8\linewidth}
\includegraphics[width=6.38cm,height=4.95cm]{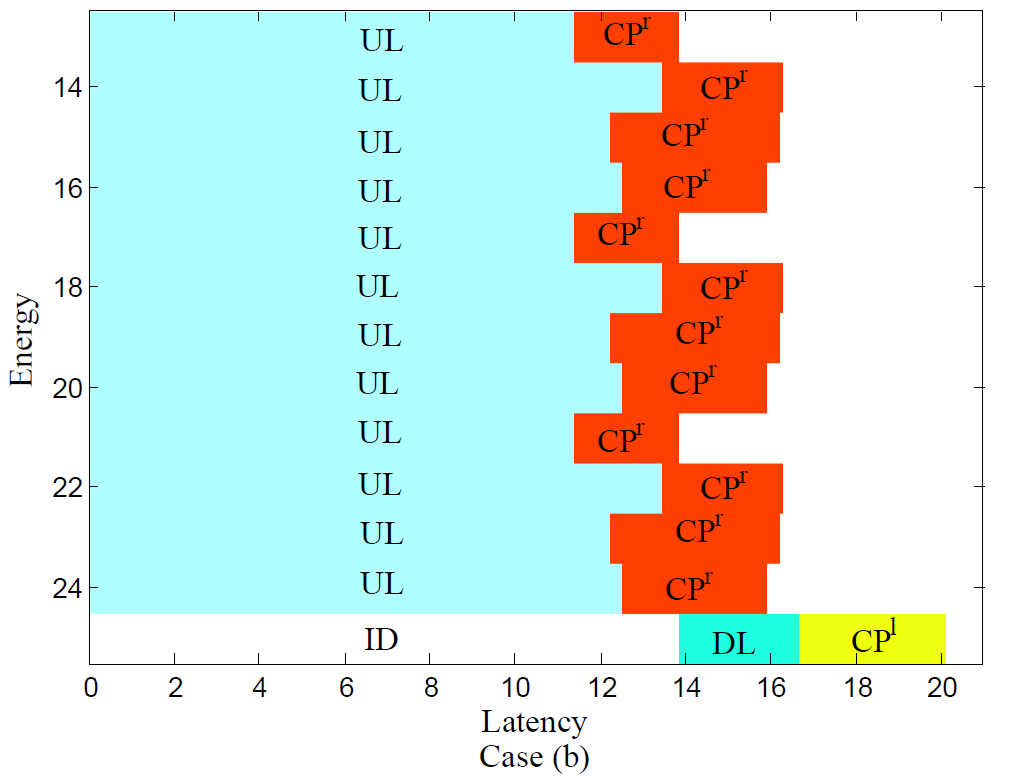}
\end{minipage}
\caption{Timeline for the parallel implementation corresponding to the optimum solution for $L=20$ s for the call graph in Fig. \ref{g_ex} (see Fig. \ref{no-gain}).}\label{time}
\vspace{-1.5em}
\end{figure}
Following similar consideration, if $X_n(k)=\mathrm{DL}$, the state of the task node $\mathrm{T}_n$ can be updated as
%task $\mathrm{T}_n$ is downloading output bits of its parent node $\mathrm{T}_m$, and we have $[b^{dl}_{m,n}(k+1)=b^{dl}_{m,n}(k)-(\log_2(1+(2^{C^{dl}}-1)N^{dl}(k))/N^{dl}(k))\epsilon]^+$ for all $m\in\mathcal{P}(n)$ with $I_m=1$.  At the end of the corresponding time slot the task node $\mathrm{T}_n$ goes into state $\mathrm{CP^l}$ if $b^{dl}_{m,n}(k+1)=0$ and $X_m(k)=\mathrm{CM}$ for all $m\in\mathcal{P}(n)$. Therefore,
\begin{equation}\label{stat1}
X_n(k+1)=\left\{\begin{array}{ll}
\mathrm{DL} & \mathrm{\textbf{if}~} b^{dl}_{m,n}(k+1)>0~\mathrm{for ~any }~m\in\mathcal{P}(n)\\
\mathrm{CP^l} &\mathrm{\textbf{if}~} b^{dl}_{m,n}(k+1)=0~\mathrm{and}~ X_m(k)=\mathrm{CM}\\
&\mathrm{for ~all }~m\in\mathcal{P}(n)
\end{array}\right..
\end{equation}
Moreover, if $X_n(k)=\mathrm{CP^l}$, we have
\begin{equation}\label{stat3}
X_n(k+1)=\left\{\begin{array}{ll}
\mathrm{CP^l}& \mathrm{\textbf{if}~} c^l_n(k+1)>0\\
\mathrm{UL}& \mathrm{\textbf{if}~} I_{\mathcal{C}(n)}=1 ~\mathrm{and}\\
&c^l_n(k+1)=0~\mathrm{and~}n\in\mathcal{V}\backslash\mathcal{V}_\mathrm{D}\\
\mathrm{CM}&\mathrm{otherwise}
%\mathrm{CM}& \mathrm{\textbf{if}~} I_{\mathcal{C}(n)}=1 ~\mathrm{and}\\ &c^l_n(k+1)=0~\mathrm{and~}n\in\mathcal{V}_\mathrm{D}\\
%\mathrm{CM}& \mathrm{\textbf{if}~} I_{\mathcal{C}(n)}=0 ~\mathrm{and}~ c^l_n(k+1)=0\\
%\mathrm{CM}& \mathrm{\textbf{if}~} n  \mathrm{~is ~the~ root ~node~ and}~ c^l_n(k+1)=0
\end{array}\right.,
\end{equation}
and, if $X_n(k)=\mathrm{CP^r}$, we can write
%the nodes goes to state $\mathrm{CM}$ after its remote processing is finished and hence the update equation of this node is given by
\begin{equation}\label{stat4}
X_n(k+1)=\left\{\begin{array}{ll}
\mathrm{CP^r}&\mathrm{\textbf{if}} ~c^r_n(k+1)>0\\
\mathrm{CM}&\mathrm{\textbf{if}} ~c^r_n(k+1)=0
\end{array}\right.,
\end{equation}
where $c^r_n(k+1)$ is calculated as $c^r_n(k+1)=[c^r_n(k)-(f^r/N^{r}(k))\epsilon]^+$. If $X_n(k)=\mathrm{CM}$, we always have $X_n(k+1)=\mathrm{CM}$ and,
if $X_n(k)=\mathrm{ID}$, we have
\begin{equation}\label{stat5}
X_n(k+1)=\left\{\begin{array}{ll}
\mathrm{DL}&\mathrm{\textbf{if}} ~I_n=0 ~\mathrm{and} ~ I_m=1 ~\mathrm{for~ some}\\
&m\in\mathcal{P}(n)\mathrm{ ~with  }  ~X_m(k)=\mathrm{CM}\\
\mathrm{UL}&\mathrm{\textbf{if}} ~I_n=1~\mathrm{and}  ~X_m(k)=\mathrm{CM} ~\mathrm{for~ all}\\
& m\in\mathcal{P}(n) \mathrm{~and~} m\in\mathcal{V}_\mathrm{D}\\
\mathrm{CP^l}&\mathrm{\textbf{if}} ~I_n=0~\mathrm{and}~ I_m=0 ~\mathrm{for ~all~}m\in\mathcal{P}(n)\\
&\mathrm{with}  ~X_m(k)=\mathrm{CM}\\
\mathrm{CP^r}&\mathrm{\textbf{if}} ~I_n=1~\mathrm{and}  ~X_m(k)=\mathrm{CM} ~\mathrm{for~ all}\\
& m\in\mathcal{P}(n) \mathrm{~and~} m\in\mathcal{V} \backslash \mathcal{V}_\mathrm{D}\\
\mathrm{ID}&\mathrm{otherwise}
\end{array}\right..
\end{equation}

Based on the discussion above, the values $N^l(k)$, $N^r(k)$, $N^{ul}(k)$ and $N^{dl}(k)$ are calculated at each time $t_k$ according to the states of nodes as
$N^{ul}(k)=\sum_{n=1}^{|\mathcal{V}|}  \mathrm{1}(X_n(k)=\mathrm{UL})$, $N^{l}(k)=\sum_{n=1}^{|\mathcal{V}|}\mathrm{1}(X_n(k)=\mathrm{CP^l})$, $N^{r}(k)=\sum_{n=1}^{|\mathcal{V}|}\mathrm{1}(X_n(k)=\mathrm{CP^r})$ and $N^{dl}(k)=\sum_{n=1}^{|\mathcal{V}|} \sum_{m\in\mathcal{P}(n)} \mathrm{1}(X_n(k)=\mathrm{DL}~\mathrm{and} ~b^{dl}_{m,n}(k)>0 ~\mathrm{and}~ X_m(k)=\mathrm{CM}$,
where $\mathrm{1}(\cdot)$ is the indicator function.

\begin{figure}
\centering
\includegraphics[scale=.3]{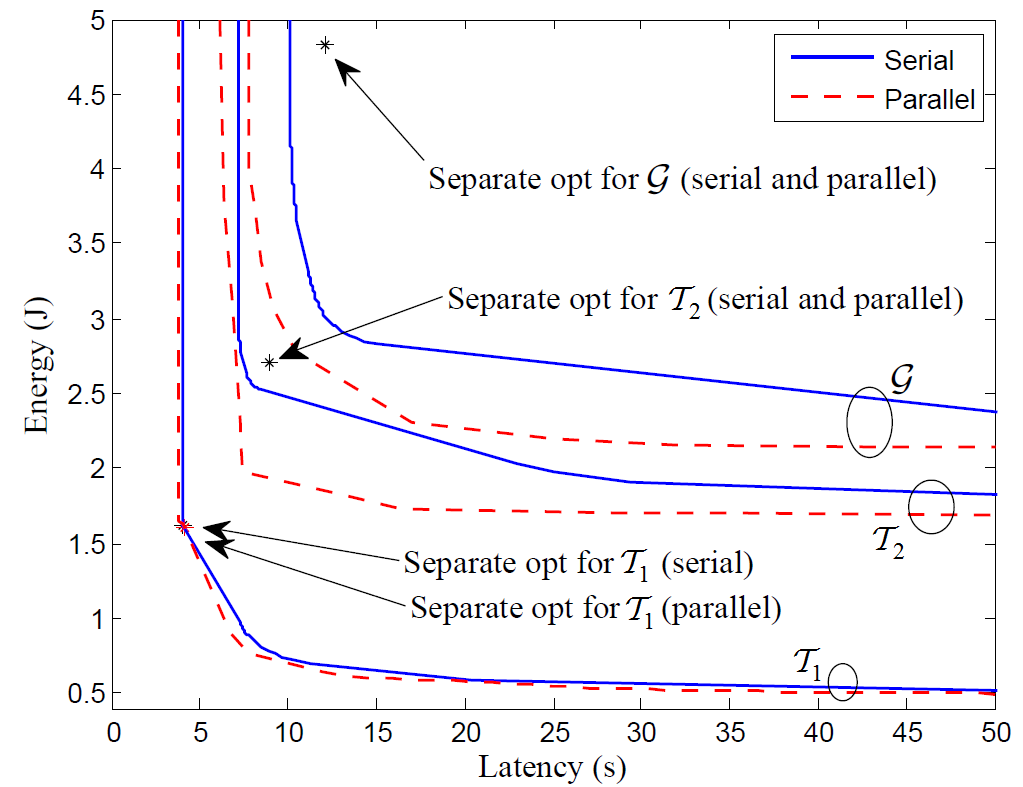}
\caption{Energy and latency trade-off for call graph $\mathcal{G}$ in Fig. \ref{graph2} and the subtrees $\mathcal{T}_1$ and $\mathcal{T}_2$ with $v_1=0$, $v_2=v_4=v_{12}=0.6\times 10^9$, $v_3=0.24\times 10^9$, $v_5=0.4\times 10^9$, $v_6=v_9=v_{14}=2\times 10^9$, $v_7=v_8=1.1\times 10^9$, $v_{10}=0.66\times 10^9$, $v_{11}=v_{13}=1\times 10^9$, $v_{15}=0.2\times 10^9$  CPU cycles, $b_{1,2}=b_{3,5}=b_{3,6}=b_{5,10}=b_{9,12}=b_{11,14}=b_{12,14}=5\times10^6$, $b_{2,3}=15\times10^6$, $b_{2,4}=9.7\times10^6$, $b_{4,7}=b_{4,8}=8.5\times10^6$, $b_{4,9}=3\times10^6$, $b_{6,10}=8\times10^6$, $b_{7,11}=b_{8,11}=1.2\times10^6$, $b_{10,13}=b_{13,15}=10\times10^6$ and $b_{14,15}=15.5\times10^6$ bits.}\label{dep}
\vspace{-1.5em}
%For the serial implementation, the cost to run this program remotely is equal to $E=0.85$ (J) with the minimum latency of  $L=159$ (sec)
\end{figure}
Finally, at the end of each time interval $(t_k,t_{k+1}]$ the energy consumed by the mobile is updated as
\begin{equation}\label{ene_end}
\begin{split}
E(k+1)&=E(k)\\
&+\sum_{n\in\mathcal{V}}\sum_{m\in\mathcal{P}(n)}\mathrm{1}\left(X_n(k)=\mathrm{DL}~\mathrm{and}\right.\\
&\left.b^{dl}_{m,n}(k)>0 ~\mathrm{and}~ X_m(k)=\mathrm{CM}\right)\\
&(P^{rx}+P^{rf})\epsilon\\
&+\sum_{n\in\mathcal{V}}\mathrm{1}\left(X_n(k)=\mathrm{UL}\right)(\bar{P}_{n,\mathcal{C}(n)}+P^{rf})\epsilon\\
&+\sum_{n\in\mathcal{V}}\mathrm{1}\left(X_n(k)=\mathrm{CP^l}\right)\frac{P^l}{N^{l}(k)}\epsilon.
\end{split}
\end{equation}
The latency is instead given by the smallest value $t_k$ such that $X_{|\mathcal{V}|}(k)=\mathrm{CM}$ for the root node $\mathrm{T}_{|\mathcal{V}|}$. We observe that (\ref{ene_end}) assumes that transmissions and computations last for the period of duration $\epsilon_d$ even if the task completed at some time within the interval. This implies that (\ref{ene_end}) and the corresponding latency are upper bounds on the actual energy and latency that become increasingly tight as $\epsilon_d$ become smaller.

\bibliographystyle{IEEEtran}
\bibliography{refrences2}

\end{document}